\documentclass[12pt]{article}
\usepackage{amsmath, amssymb}
\usepackage{amsfonts}
\usepackage{graphicx,psfrag,epsf}
\usepackage{enumerate}
\usepackage{natbib, xcolor}
\usepackage{url} % not crucial - just used below for the URL 
\usepackage[noend]{algpseudocode}
\usepackage{algorithm}
\usepackage{booktabs}
\usepackage{multirow}
\usepackage{verbatim}
\usepackage{array}
\usepackage{setspace}
\usepackage{authblk}
\usepackage[compact]{titlesec}
\newcommand{\blind}{0}

\newcommand{\bdelta}{\mbox{\boldmath $\delta$}}

\newcolumntype{L}[1]{>{\raggedright\let\newline\\\arraybackslash\hspace{0pt}}m{#1}}
\newcolumntype{C}[1]{>{\centering\let\newline\\\arraybackslash\hspace{0pt}}m{#1}}
\newcolumntype{R}[1]{>{\raggedleft\let\newline\\\arraybackslash\hspace{0pt}}m{#1}}
\algnewcommand\And{\textbf{and}}

\begin{document}

\def\spacingset#1{\renewcommand{\baselinestretch}%
{#1}\small\normalsize} \spacingset{1}

%%%%%%%%%%%%%%%%%%%%%%%%%%%%%%%%%%%%%%%%%%%%%%%%%%%%%%%%%%%%%%%%%%%%%%%%%%%%%%

\if0\blind
{

  \title{\bf Quantifying uncertainty of individualized treatment effects in right-censored survival data: A comparison of Bayesian additive regression trees and causal survival forest}
\author[1]{Daijiro Kabata}
\author[2]{Nicholas C. Henderson}
\author[3,4,*]{Ravi Varadhan}
\affil[1]{\small Center for Mathematical and Data Science, Kobe University, Kobe, Hyogo, Japan}
\affil[2]{\small Department of Biostatistics, University of Michigan, Ann Arbor, Michigan, USA}
\affil[3]{\small Department of Oncology, School of Medicine, Johns Hopkins University, Baltimore, Maryland, USA}
\affil[4]{\small Department of Biostatistics, Johns Hopkins University, Baltimore, Maryland, USA}
\affil[*]{{\small corresponding author

\textit{email:} ravi.varadhan@jhu.edu}}

\date{}
  \maketitle
} \fi

\if1\blind
{
  \bigskip
  \bigskip
  \bigskip
  \begin{center}
    {\LARGE\bf Bayesian additive regression trees and honest causal forest for uncertainty quantification of individualized treatment effects in observational, right-censored data: A comparative study}
\end{center}
  \medskip
} \fi

\bigskip
\begin{abstract}
Estimation of individualized treatment effects (ITE), also known as conditional average treatment effects (CATE), is an active area of methodology development. However, much less attention has been paid to the quantification of uncertainty of ITE/CATE estimates in right-censored survival data. Here we undertake an extensive simulation study to examine the coverage of interval estimates from two popular estimation algorithms, Bayesian additive regression trees (BART) and causal survival forest (CSF). We conducted simulation designs from 3 different settings: first, in a setting where BART was developed for an accelerated failure time model; second, where CSF was developed; and finally, a ``neutral'' simulation taken from a setting where neither BART nor CSF was developed. 
BART outperformed CSF in all three simulation settings. Both the BART and CSF algorithms involve multiple hyperparameters, and BART credible intervals had better coverage than the CSF confidence intervals under the default values, as well as under optimized values, of these hyperparameters.
\end{abstract}

\noindent%
{\it Keywords:} causal inference; coverage evaluation; interval estimation; heterogeneity of treatment effect; machine learning; observational studies; precision medicine
\vfill

\newpage
\spacingset{1.45}

\section{Introduction}
\label{sec:intro}
Estimating individual treatment effects (ITE) in survival analysis is an increasingly important goal in health research. Traditional approaches such as Cox regression with predefined interactions or subgroup analyses can be limited by linearity assumptions and the requirement to specify candidate effect modifiers in advance. Over the past decade, machine-learning methods have gained prominence for flexibly modeling survival data in observational studies, where confounding and complex relationships between covariates and outcomes can obscure which patients benefit most from a particular intervention.

Bayesian Additive Regression Trees (BART) has emerged as a prominent nonparametric approach for assessing individualized treatment effects in both continuous and time-to-event settings (\cite{hill2011, hahn2020}). In survival analysis, semiparametric BART implementations often adopt an accelerated failure time (AFT) perspective, modeling log survival times as sums of regression trees while relying on Bayesian priors to regularize the fits (\cite{henderson2020individualized}; \cite{sparapani2016}). Sparapani et al. have stated that BART’s posterior intervals in survival contexts achieve or slightly exceed nominal coverage levels, an aspect that underscores the method’s reputation for robust uncertainty quantification. However, BART’s intervals are sometimes more conservative, potentially being wider than necessary if the model overestimates variability.

In parallel to the development of BART-based methods, other tree-based ensemble methods for causal inference (\cite{athey2016}) have also been adapted to right-censored data. One such recently developed method is Causal Survival Forests (CSF) (\cite{cui2023}). By extending the generalized random forest framework to a survival setting, CSF identifies which covariates modify treatment effects using data-driven partitioning, and it constructs confidence intervals through orthogonalization and resampling. Existing work on CSF has indicated that its interval estimates can be well-calibrated given adequate sample sizes, a sufficient number of trees, and appropriate tuning parameters (\cite{cui2023}). Nevertheless, such coverage studies have typically involved settings where BART-based survival models were not simultaneously examined, making it difficult to draw direct comparisons of how often each method’s 95\% uncertainty interval contains the true individualized treatment effect (ITE).

Recent reviews and simulation studies have underscored that although both BART and forest-based approaches exhibit encouraging performance for uncovering HTE, head-to-head evaluations of interval coverage in survival analyses remain scarce (Caron et al., 2022). While some articles have noted that BART tends to provide slightly more conservative uncertainty intervals, and that CSF can match nominal coverage if implemented with sufficiently large ensembles, none appear to have systematically compared the two under identical survival data-generating processes. The lack of direct coverage-focused comparisons has left open questions regarding which method is more reliable when investigators require high-confidence interval estimates of individualized effects. These questions are pertinent not only in theory but also in practical applications where confidence intervals inform clinical or policy decisions about patient-specific benefits and harms.

In this paper, we address this gap by examining the 95\% coverage of BART and CSF in a series of extensive simulation studies motivated by three prior studies: \cite{henderson2020individualized}, \cite{cui2023}, and \cite{hu2021}. The first scenario incorporates an accelerated failure time model similar to that proposed for BART, the second employs discrete threshold structures favoring forest-based splitting, and the third reflects a more neutral Weibull-based design. By embedding these simulations in an observational framework with moderate to strong confounding, we replicate circumstances commonly faced in real-world data analyses. Although previous research has highlighted BART’s strong performance in interval coverage and CSF’s value for identifying subgroup effects, a direct comparison of the methods under consistent conditions has been lacking. Our findings contribute novel evidence about how well each method’s intervals capture true heterogeneity in survival settings, shedding light on the circumstances under which researchers might prefer one approach over the other for achieving valid inference on individualized treatment effects.

% DK added 250317
\section{Background and Methods Description}

\subsection{Individualized Treatment Effects in Survival Analysis}
Our interest is in settings where one wants to evaluate the effect
of a treatment on an endpoint that is a time $T$ to an event of interest. 
In biomedical studies, one often cannot directly observe $T$, but can
instead observe the time of last follow-up $Y$. This equals either a censoring time $C$ if the last follow-up  
time occurs before $T$ and equals $T$ otherwise.
That is, $Y$ is the minimum of $T$ and $C$: $Y = \min\{ T, C \}$.
The binary variable $\delta = I(T < C)$ equals $1$ if the time to the event
of interest was observed and equals $0$ otherwise.
The variable $A$ indicates the treatment group membership with $A = 0$
denoting membership in the control group of the study and $A = 1$ denoting membership in the
active treatment arm of the study. In addition to $(Y, \delta, A)$,  we 
observe $\mathbf{X} = (x_{1}, \ldots, x_{p})$ which
is a collection of $p$ pre-treatment covariates.
The data collected in such settings consist of $n$ independent measurements of the 
form $\{ Y_{i}, \delta_{i}, A_{i}, \mathbf{X}_{i} \}$,
where $i = 1, \ldots, n$.

To characterize HTE in settings with survival outcomes, one first needs to specify the 
treatment effect of interest. To define treatment effects, it is helpful to first define survival counterfactual (or potential) outcomes $T(1)$ and $T(0)$. Here, $T(a)$ represents the survival time of an individual in treatment group $a$, and
counterfactual survival outcomes $T(a)$ and the survival outcome $T$ are connected by the following
\begin{equation}
T = AT(1) + (1 - A)T(0). \nonumber 
\end{equation}
Frequently, a treatment effect of interest can be expressed as the expected difference in transformed counterfactual survival outcomes $b(T(1))$ and $b(T(0))$, for a monotone function of interest $b(.)$. For example, one might choose $b(x) = I(x > t)$ for t-year
survival probabilities or $b(x) = \min\{x, \tau\}$ for restricted mean survival time (\cite{royston2013}). The expected difference $E[b(T(1))] - E[b(T(0))]$ represents an average treatment effect because it does not condition on individual patient characteristics in any way.

Individualized treatment effects (ITE) -- which are often referred to as conditional average treatment effects (CATEs) -- represent average treatment effects among individuals that share a common covariate vector. Specifically, for a transformation function of interest $b$, the ITE at covariate vector $\mathbf{x}$ is defined as
\begin{equation}
E[ b(T(1)) \mid \mathbf{X} = \mathbf{x}] - E[b(T(0)) \mid \mathbf{X}=\mathbf{x}]. \nonumber 
\end{equation}
Our main focus in this paper is on the ITE that contrasts expected survival on the log-time scale. This ITE function $\theta(\mathbf{x})$ is defined as
\begin{equation}
\theta(\mathbf{x}) = \mathbb{E}[\log(T(1)) \mid \mathbf{X}=\mathbf{x}] \;-\; \mathbb{E}[\log (T(0)) \mid \mathbf{X}=\mathbf{x}]. \nonumber 
\end{equation}
Under standard causal assumptions (e.g.\ consistency and no unmeasured confounding), this quantity corresponds to a contrast of the conditional expectation of the log of $T$
with and without a treatment $A\in{\{0,1\}}$ for a patient with covariates $\mathbf{x}$. That is,  
\begin{equation}\label{eq:ite}
\theta(\mathbf{x}) \;=\; \mathbb{E}[\log (T)\mid A=1, \mathbf{X}=\mathbf{x}] \;-\; \mathbb{E}[\log (T)\mid A=0, \mathbf{X}=\mathbf{x}],
\end{equation}
In essence, $\theta(\mathbf{x})$ captures heterogeneity in treatment effects by quantifying how much longer (or shorter) an individual is expected to survive, on the log scale, with the treatment compared to without. We next describe two flexible modeling approaches for estimating $\theta(\mathbf{x})$ from censored survival data: AFT-BART and CSF estimators.

\subsection{AFT-BART}
The BART-AFT method described in \cite{henderson2020individualized} focuses 
on utilizing flexible Bayesian machine learning tools to model individual-level survival distributions.
One can then estimate the ITE function of interest by noting that any ITE
is a feature of this full survival distribution. 
Individual-level survival distributions are built by using the Bayesian Additive Regression Trees (BART) framework 
(\cite{chipman2010}) to flexibly model covariate and treatment effects on the log failure-time scale. Specifically, the model posits that
\begin{equation}
\log(T) \;=\; m(A,\mathbf{X}) \;+\; W, \label{eq:np_aft_model}
\end{equation}
where \(m(A,\mathbf{X})\) is referred to as the regression function and \(W\) is a residual term with mean zero. Rather than assuming a linear (or otherwise simple) form for \(m\), AFT-BART uses sums of regression trees:
\[
m(A,\mathbf{x}) \;=\; \sum_{j=1}^{J} g(A,\mathbf{X}; \mathcal{T}_j, \mathcal{B}_j),
\]
where each \(g(\cdot)\) is a small tree with splitting rules \(\mathcal{T}_j\) and leaf parameters \(\mathcal{B}_j\). A regularization prior on these trees helps to prevent overfitting by shrinking individual tree contributions.

Under AFT model (\ref{eq:np_aft_model}), the ITE function $\theta(\mathbf{x})$ for the difference in log-survival
can be expressed as the difference in the regression function as:
\begin{eqnarray}
\theta(\mathbf{x}) &=& E\{ \log T_{i} \mid A = 1, \mathbf{X} = \mathbf{x} \} 
- E\{ \log T \mid A = 0, \mathbf{X} = \mathbf{x} \} \nonumber \\
&=& m(1, \mathbf{x}) - m(0, \mathbf{x}). \nonumber 
\end{eqnarray}

%One can also define a CATE function on an alternative scale. For example, 
%\begin{equation}
%\xi(\mathbf{x}) = E\{ T_{i} \mid A_{i} = 1, \mathbf{x}_{i} = \mathbf{x} \} 
%- E\{ T_{i} \mid A_{i} = 0, \mathbf{x}_{i} = \mathbf{x} \}. \nonumber 
%\end{equation}

\cite{henderson2020individualized} introduced two versions of AFT-BART: a fully nonparametric (AFT-BART-NP) and a semiparametric (AFT-BART-SP) specification. The main distinction between the NP and SP versions lies in how the residual distribution is modeled. The fully nonparametric version (AFT-BART-NP) places a centered Dirichlet process mixture model prior on the error distribution, allowing maximum flexibility. In contrast, the semiparametric version (AFT-BART-SP) assumes that the residual follows a Gaussian distribution with mean zero while still retaining the flexible BART prior for \(m(a,\mathbf{x})\).

In this work, we primarily evaluate the semiparametric version, AFT-BART-SP, for two reasons. First, by incorporating some parametric structure on the residuals, AFT-BART-SP can improve efficiency and interpretability without sacrificing much flexibility in modeling covariate or treatment effects. Second, we are comparing performance against a causal survival forest (CSF), which likewise adopts certain structural assumptions on survival outcomes. Using the SP variant thus provides a more directly comparable framework than the fully nonparametric specification.

\subsection{Causal Survival Forest}
The CSF is a tree-based ensemble method for estimating heterogeneous treatment effects in right-censored survival data \citep{cui2023}. It extends the generalized random forest framework to a survival setting, focusing on a difference in survival metrics. Under assumptions of no unmeasured confounding and independent censoring, $\theta(\mathbf{x})$ is identifiable from observed data. 
CSF grows a large number of survival trees on bootstrap samples \citep{breiman2001}, using splitting rules that maximize heterogeneity in treatment effects while accounting for censoring via survival-specific criteria (e.g.\ log-rank statistics). Each tree partitions the covariate space; within each leaf, a leaf-specific treatment effect is estimated. The forest-wide estimate is the average across all trees. This ensemble approach is robust, capturing complex interactions and reducing variance.
CSF employs sample splitting (“honest trees”) and orthogonalization for valid inference \citep{wager2018}. Specifically, the algorithm splits data to separate the process of finding splits from the process of estimating treatment effects in each leaf. Orthogonalization (double robustness) uses nuisance estimates (e.g.\ propensity scores, censoring hazards) to form a de-biased pseudo-outcome before tree building.

\section{Simulation Design and Evaluation Metrics} 

To comprehensively compare the performance of individualized treatment effect (ITE) estimation for survival times using AFT-BART-SP and CSF across various scenarios, we conducted simulation studies based on the designs of \cite{henderson2020individualized}, \cite{cui2023}, and \cite{hu2021}. Since \cite{henderson2020individualized} originally introduced the AFT-BART framework, their simulation setting may be inherently more favorable to BART; likewise, the setting proposed by \cite{cui2023} — which introduced CSF — may favor CSF. Therefore, in order to make the comparison more balanced, we also adopted the design of \cite{hu2021}.

For ITE estimation using AFT-BART-SP, we employed the \texttt{IndivAFT} function from the \texttt{AFTree} R package (\cite{hendersonaft}). 
For CSF, all ITE estimates and uncertainty intervals were generated using the \texttt{causal\_survival\_forest} function
from the \texttt{grf} R package (\cite{tibshirani2018}). In each case, we considered two types of models: one with default parameter settings and another with reduced regularization (hereafter referred to as ``improved default''). Specifically, we changed the shrinkage parameter \texttt{min.node.size} from the default of 5 to 2 in \texttt{AFTree::IndivAFT}, and the parameter \texttt{k} from 2 to 1 in \texttt{grf::causal\_survival\_forest}. 
To improve estimation performance, we considered the estimated propensity score as an additional covariate when fitting both the AFT-BART and CSF models, as recommended in previous studies (\cite{athey2016, wager2018}). To estimate propensity scores, we employed a Super Learner-based ensemble approach combining three prediction algorithms: a generalized linear model, a generalized additive model, and a random forest. The weights used to form this ensemble were determined by minimizing the weighted standardized mean differences of covariates between treatment groups. Specifically, we implemented this estimation using the \texttt{weightit} function in the \texttt{WeightIt} R package. 

\begin{figure}[H]
        \caption{Results of the Simulation Study based on Henderson et al. (2020)}
        \includegraphics[width=17cm]{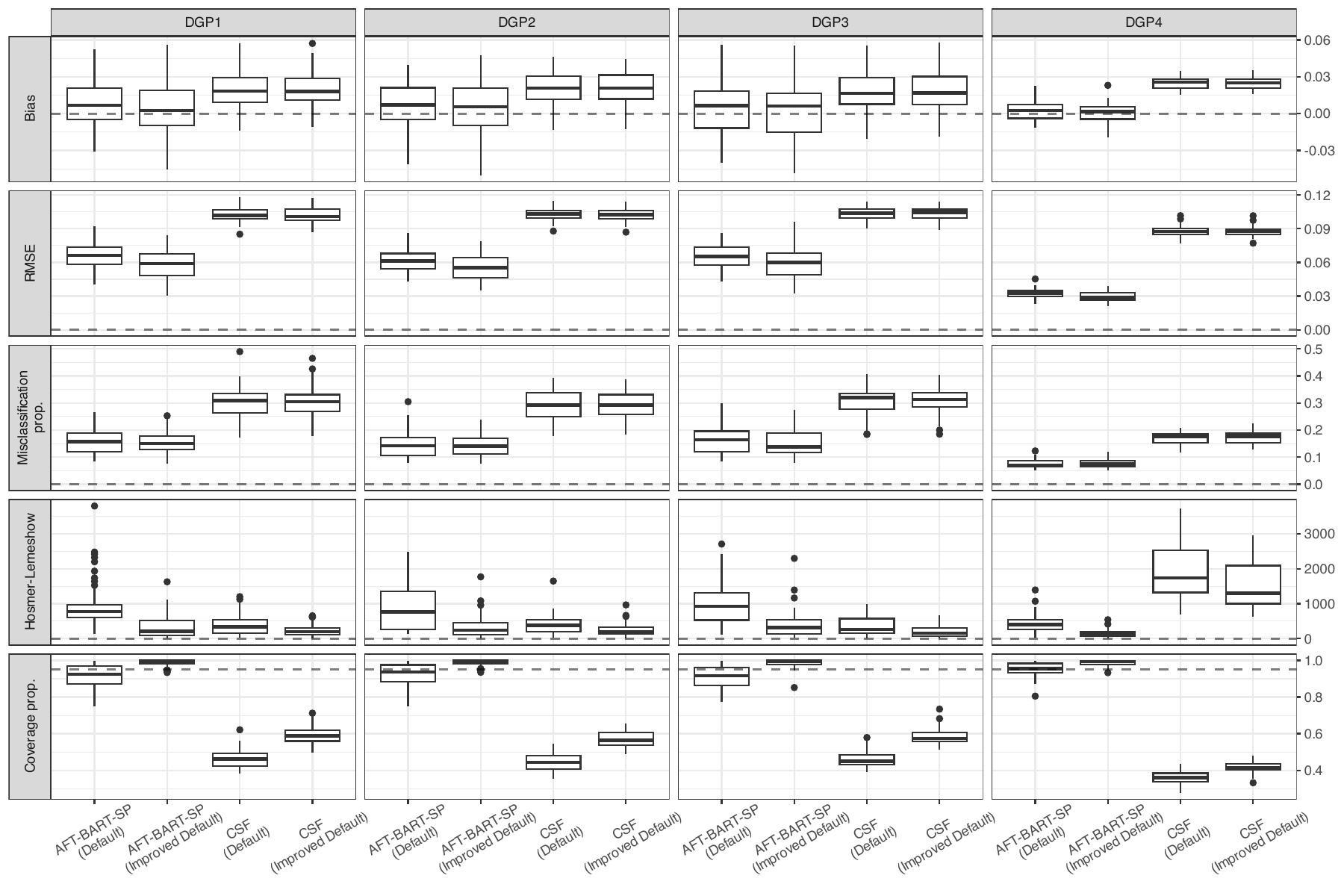}        
\end{figure}

In all our simulations, the sample size was fixed at $n = 1,000$ individuals, and each setting was replicated 50 times to evaluate performance under repeated sampling. Across the replications, we assessed the performance of ITE estimation using five main metrics: bias, root mean-squared error (RMSE), misclassification proportion, Hosmer--Lemeshow statistic, and average coverage of 95\% intervals. 
The misclassification proportion was defined as the proportion of individuals assigned to the incorrect treatment category based on discrepancies between the estimated and true signs of the ITE. Specifically, for the AFT-BART-SP model, misclassification was defined as cases where the true sign of \(\theta(\mathbf{x}_i)\) is positive but the posterior probability of the ITE having a negative sign exceeds 0.5, or cases where the true sign is negative but the posterior probability of the ITE having a positive sign exceeds 0.5. For the CSF model, misclassification occurred when the estimated sign of \(\hat{\theta}(\mathbf{x}_i)\) differed from the true sign of \(\theta(\mathbf{x}_i)\). Furthermore, average coverage was calculated as the proportion of times the 95\% credible intervals from the AFT-BART-SP model and the 95\% confidence intervals from the CSF model contained the true value of \(\theta(\mathbf{x}_i)\).

\subsection{Setting Following Henderson et al.\ (2020)}

In this simulation framework, based on \cite{henderson2020individualized}, we used a dataset consisting of 1{,}000 individuals randomly sampled from the \texttt{AFTree::solvd\_sim} dataset available in the \texttt{AFTree} package. 
The \texttt{AFTree::solvd\_sim} dataset is a synthetic dataset that includes variables similar to those collected in the Studies of Left Ventricular Dysfunction (SOLVD) Treatment and Prevention Trials (\cite{solvd1991}).

To simulate survival outcomes, we first estimated the regression function $\tilde{m}(A,\mathbf{x})$ for $A \in \{0,1\}$ using the nonparametric AFT-BART model.
Survival times $T$ were then generated as
\[
\log(T) = A \tilde{m}(1,\mathbf{x}) + (1 - A)\tilde{m}(0,\mathbf{x}) + W.
\]
The regression function, treatment assignment variables, and covariate vectors were held constant across all simulation replications. New values of the error terms $W$ were generated randomly for each simulation replication. Four different data-generating processes (DGP) were considered for the distribution of the residual term: (DGP1) Gaussian distribution, (DGP2) Gumbel distribution with mean zero, (DGP3) standardized Gamma distribution with mean zero (Std-Gamma), and (DGP4) a mixture of three $t$-distributions (each with three degrees of freedom, referred to as ``T-mixture''). Independent right-censoring was generated from a $\mathrm{Uniform}(6.5,10)$ distribution, introducing an overall censoring rate of approximately 15\%.

\subsection{Setting Following Cui et al. (2023)}
Following the simulation design of \cite{cui2023}, we generated \(p=15\) covariates \(\mathbf{X} = (X_1, \dots, X_{15})\) from correlated \(\mathrm{Uniform}\) distributions, where correlation was induced via a covariance matrix $\mathbf{V}$ whose \((i,j)\)th entry is \(0.5^{\lvert i-j\rvert}\). Specifically, $\mathbf{X}$ was generated by setting $\mathbf{X} = \mathbf{L}\mathbf{U}$, where $\mathbf{V} = \mathbf{L}\mathbf{L}^{T}$ is the Cholesky factorization of $\mathbf{V}$
and where $\mathbf{U}$ is a vector containing independent \(\mathrm{Uniform}(0,1)\) random variables.

The treatment indicator \(A\) was drawn from a Bernoulli distribution with treatment assignment probability \(e(\mathbf{X}) = P(A=1 \mid \mathbf{X})\). Counterfactual survival times \(T\) and censoring times \(C\) were generated using AFT models, Cox proportional hazards models, Poisson distributions, or Uniform distributions, depending on the specific DGP, as detailed below:
\begin{flalign*}
\text{DGP1: }&e(\mathbf{X}) = (1+\beta(X_1;2,4))/4, && \\
&\log(T) = -1.85 -0.8I(X_1<0.5) +0.7X_2^{1/2} +0.2X_3 + \left(0.7 -0.4I(X_1<0.5) -0.4X_2^{1/2}\right)A + \varepsilon, && \\
&\lambda_C(t|A,X) = \lambda_0(t)\exp\left[f(A,\mathbf{X})\right], && \\
&\text{where } \varepsilon \sim \mathrm{N}(0,1), \lambda_0(t)=2t, \\
&\quad\quad\quad f(A,\mathbf{X}) = -1.75 -0.5X_2^{1/2} +0.2X_3 + \left(1.15 +0.5I(X_1<0.5)-0.3X_2^{1/2}\right)A, \\
&\quad\quad\quad \text{maximum follow-up is }1.5; &&
\end{flalign*}
\begin{flalign*}
\text{DGP2: }&e(\mathbf{X}) = (1 + \beta(X_2;2,4))/4, &&\\
&\lambda_T(t|A,X) = \lambda_0(t)\exp\left[f(A,\textbf{X})\right], && \\
&C \sim \mathrm{U}(0,3), && \\
&\text{where } \lambda_0(t)=0.5t^{-1/2}, f(A,\textbf{X}) = X_1 + (-0.5 + X_2)A, \text{ maximum follow-up is }2; &&
\end{flalign*}
\begin{flalign*}
\text{DGP3: }&e(\mathbf{X}) = (1+\beta(X_1;2,4))/4, && \\
&T \sim \mathrm{Poisson}\left(X_2^2 + X_3 + 6 + 2(X_1^{1/2}-0.3)A\right), && \\
&C \sim \mathrm{Poisson}\left(12 + \log\left(1+\exp(X_3)\right)\right), && \\
&\text{maximum follow-up is }15;
\end{flalign*}
\begin{flalign*}
\text{DGP4: }&e(\mathbf{X}) = \{(1+\exp[-X_1])(1+\exp[-X_2])\}^{-1}, && \\
&T \sim \mathrm{Poisson}\left(X_2 + X_3 + \max(0,X_1-0.3)A\right), && \\
&C \sim \mathrm{Poisson}\left(1 + \log\left(1+\exp[X_3]\right)\right), && \\
&\text{maximum follow-up is }3. &&
\end{flalign*}
In DGP1 and DGP2, $\lambda_{C}(t \mid A, \mathbf{X})$ and $\lambda_{T}(t|A, \mathbf{X})$ refer to the censoring time and survival time hazard functions respectively.
In DGP1, DGP2, and DGP3, 
\(\beta(\cdot;a,b)\) denotes the Beta probability density function with shape parameters $a$ and $b$. 
%The counterfactual survival times and censoring times from Cox models were generated using inverse transform sampling as follows:
%\[
%T(a) = \left\{\frac{-\log(U)}{\exp[f(A=a,\mathbf{X})]}\right\}^{2}, \quad \text{where } U \sim \text{Uniform}(0,1).
%\]

\begin{figure}[H]
        \caption{Results of the Simulation Study based on Cui et al. (2023)}
        \includegraphics[width=17cm]{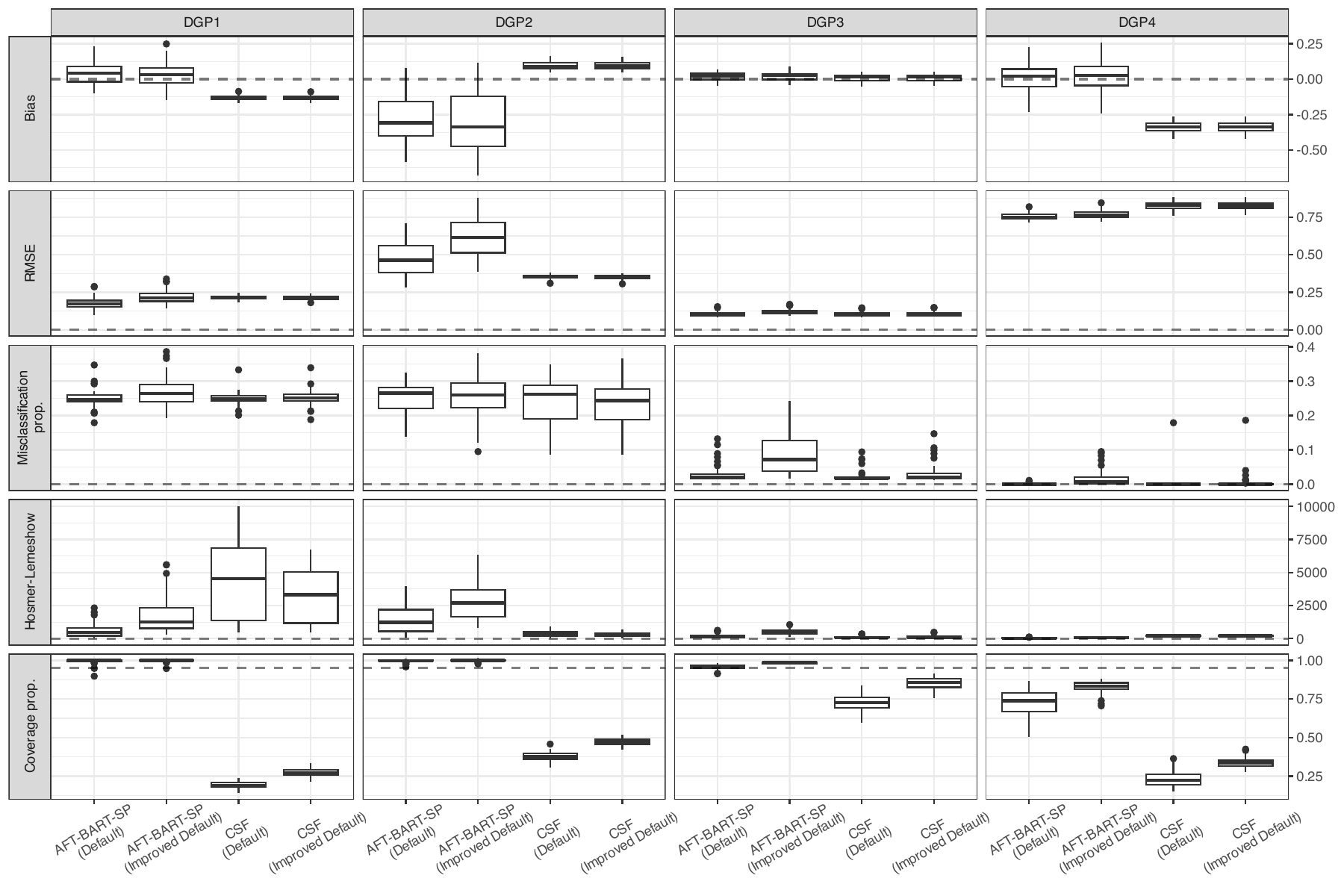}
\end{figure}

\subsection{Setting Following Hu et al.\ (2021)}
For the more ``neutral'' setting, we followed the simulation design of \cite{hu2021}. 
We generated $p=10$ independent covariates $\mathbf{X} = (X_1, \dots, X_{10})$ including five continuous variables $(X_1, \dots, X_5)$ sampled from a normal distribution with mean $0$ and standard deviation $0.35$, and the remaining five binary covariates $(X_6, \dots, X_{10})$ sampled from a Bernoulli distribution with success probability $0.5$. Based on these covariates, the treatment assignment $A$ was drawn from a Bernoulli distribution with success probability $e(\mathbf{X})$ given by, 
\[
e(\mathbf{X}) = \left\{1 + \exp[-(0.3 -0.25X_1 -2.25X_2 -0.75X_3 -0.25X_5 -0.25X_6 -0.50X_7 -1.0X_9 + 1.25X_{10})]\right\}^{-1}.
\]

To accommodate more flexible modeling of survival times, we adopted a Weibull distribution as the baseline hazard function following \cite{hu2021}. The Weibull-based Cox proportional hazards model for generating counterfactual survival times $T(a)$ is given by:
\[
\lambda_T(t \mid A=a, \mathbf{X}) = \lambda_0(t) \exp\left[ f_a(\mathbf{X}) \right], \quad \text{where}\quad \lambda_0(t)=\eta d_a t^{\eta-1}.
\]
We assumed proportional hazards by setting $\eta=2$ and specifying the treatment-group-specific scale parameter $d_a$ as $d_0=1200$ (control group) and $d_1=2000$ (treatment group). Counterfactual survival times $T(0)$ and $T(1)$ were generated using the inverse transform sampling method as follows:
\[
T(a) = \left\{\frac{-\log(U)}{d_a \exp[f_a(\mathbf{X})]}\right\}^{1/\eta}, \quad  \text{where } U \sim \text{Uniform}(0,1).
\]

We considered four treatment-effect heterogeneity scenarios (DGP1--DGP4) of increasing complexity by specifying different functional forms \(f_0(\mathbf{X})\) and \(f_1(\mathbf{X})\):
\begin{align*}
\text{DGP1: }\quad f_1(\mathbf{X}) &= -0.2 + \frac{0.1}{1 + e^{-X_1}} - 0.8\sin(X_3) - 0.1X_5^2 - 0.3X_6 - 0.2X_7, \\
f_0(\mathbf{X}) &= 0.2 - 0.5X_1 - 0.8X_3 - 1.8X_5 - 0.9X_6 - 0.1X_7; \\
\text{DGP2: }\quad f_1(\mathbf{X}) &= -0.2 + \frac{0.1}{1 + e^{-X_1}} - 0.8\sin(X_3) - 0.1X_5^2 - 0.3X_6 - 0.2X_7, \\
f_0(\mathbf{X}) &= -0.1 + 0.1X_1^2 - 0.2\sin(X_3) + \frac{0.2}{1 + e^{-X_5}} + 0.2X_6 - 0.3X_7; \\
\text{DGP3:}\quad f_1(\mathbf{X}) &= 0.5 - \frac{0.1}{1 + e^{-X_2}} + 0.1\sin(X_3) - 0.1X_4^2 + 0.2X_4 - 0.1X_5^2 \\
& + \frac{0.2}{1 + e^{-X_5}} + 0.2X_6 - 0.3X_7, \\
f_0(\mathbf{X}) &= -0.1 + 0.1X_1^2 - 0.2\sin(X_3) + \frac{0.2}{1 + e^{-X_5}} + 0.2X_6 - 0.3X_7; \\
\text{DGP4:}\quad f_1(\mathbf{X}) &= 0.5 - \frac{0.1}{1 + e^{-X_2}} + 0.1\sin(X_3) - 0.1X_4^2 + 0.2X_4 - 0.1X_5^2 - 0.3X_6, \\
f_0(\mathbf{X}) &= -0.2 + 0.5\sin(\pi X_1 X_3) + \frac{0.2}{1 + e^{-X_5}} + 0.2X_6 - 0.3X_7.
\end{align*}
The censoring times $C$ were generated independently from an exponential distribution with rate parameter 0.007 inducing a censoring rate of 20\%.

\section{Simulation Results} 

\subsection{Setting Following Henderson et al.\ (2020)}
Figure 1 summarizes results under the framework of Henderson et al.\ (2020). Overall, AFT-BART-SP outperformed CSF in terms of lower bias, RMSE, and misclassification proportion across all four DGPs. The coverage proportion for AFT-BART-SP was near or slightly below the nominal 95\% level under the default setting (around 90--95\%), but the ``improved default'' parameters substantially increased coverage closer to or even above 95\%. In contrast, CSF tended to have a substantially lower coverage (often below 60\%), indicating underestimated uncertainty. The Hosmer-Lemeshow statistic for AFT-BART-SP was sometimes large in the default setting (e.g., DGP1 or DGP3), but improved notably under the ``improved default'' configuration.

\begin{figure}[H]
        \caption{Results of the Simulation Study based on Hu et al. (2021)}
        \includegraphics[width=17cm]{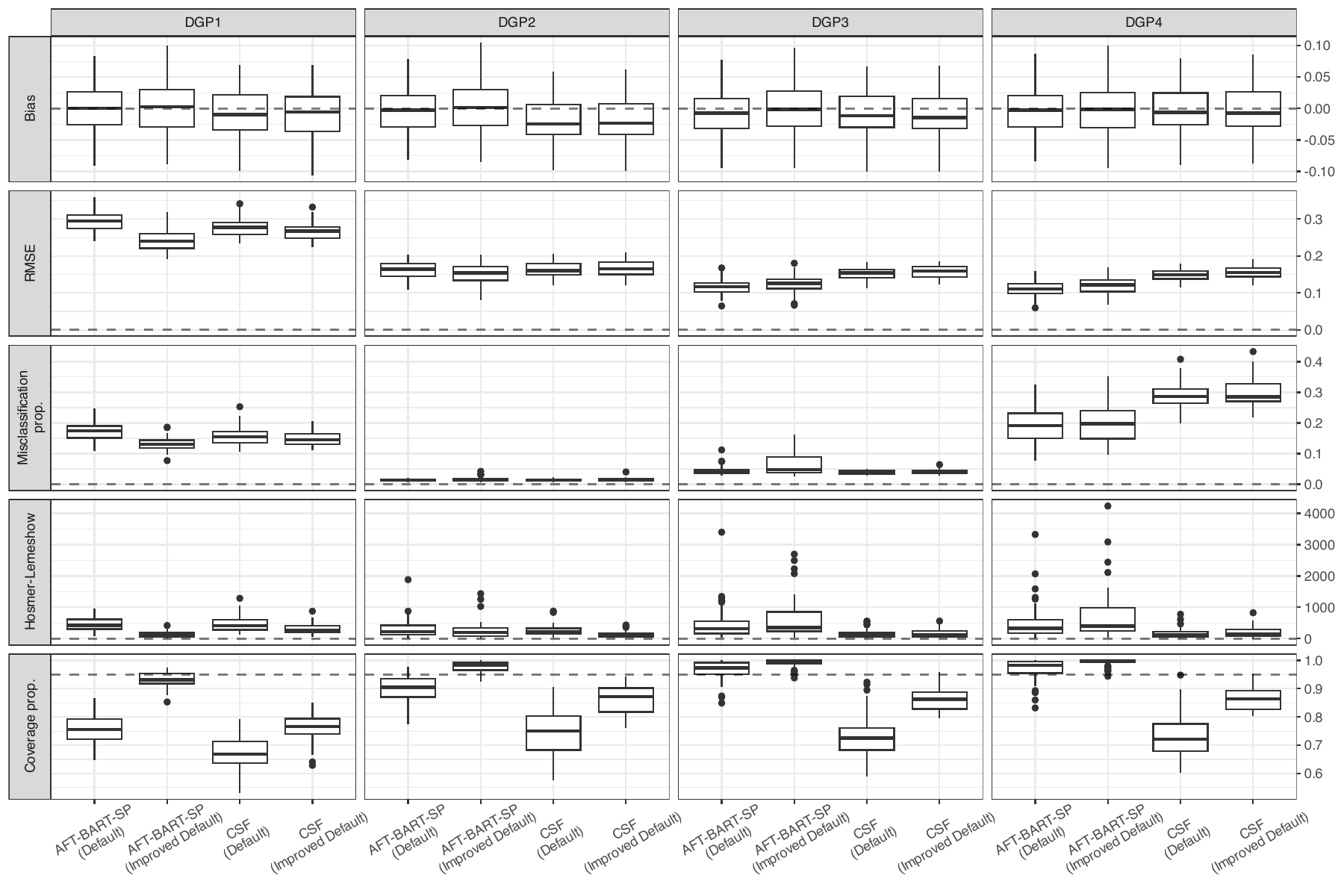}
\end{figure}

\subsection{Setting Following Cui et al.\ (2023)}
Figure 2 shows results for the designs of Cui et al.\ (2023). In this setting, CSF sometimes exhibited relatively smaller RMSE and misclassification proportion than AFT-BART-SP (e.g., DGP2). However, coverage for CSF remained much lower than 95\%, especially under the default parameters. In contrast, AFT-BART-SP achieved high coverage (above 95\%) under both default and ``improved default'' settings, although this sometimes came at the expense of increased bias or RMSE in certain DGPs (e.g., DGP2). The Hosmer-Lemeshow statistic varied considerably across scenarios; large values generally indicated calibration challenges for both methods, but especially for CSF in some cases (e.g., DGP1).

\subsection{Setting Following Hu et al.\ (2021)}
Figure 3 presents results for the Hu et al.\ (2021) framework. Across these DGPs, AFT-BART-SP again showed more stable performance, featuring smaller bias and RMSE (except for a slight increase in RMSE under the ``improved default'' in DGP3 and DGP4), as well as higher coverage proportion relative to CSF. Notably, the ``improved default'' parameters for AFT-BART-SP achieved coverage close to or above 95\% in all scenarios, whereas CSF’s coverage ranged from about 65\% to 86\%. The misclassification proportion for AFT-BART-SP also remained below or comparable to that of CSF in most DGPs, indicating that the sign of the predicted individualized effects was more accurate overall.

\medskip

In summary, across all three simulation designs, AFT-BART-SP generally exhibited more favorable estimation accuracy and more reliable coverage intervals compared to CSF, with the ``improved default'' parameters consistently enhancing calibration and coverage.

\section{Simulations under a Null HTE Setting}
In addition to the heterogeneous scenarios described above, we conducted a set of simulations under a \emph{null HTE setting}, meaning the treatment effect is the same for all individuals (i.e., there is \emph{no heterogeneity}), although the treatment effect is not necessarily zero. Our goal was to investigate whether AFT-BART-SP or CSF more accurately detects this lack of heterogeneity, focusing on how often each method \emph{incorrectly} concludes that an individual’s treatment effect differs from the overall mean.

\begin{figure}[H]
        \caption{Results of Simulation Study with the Null HTE setting.}
        \includegraphics[width=17cm]{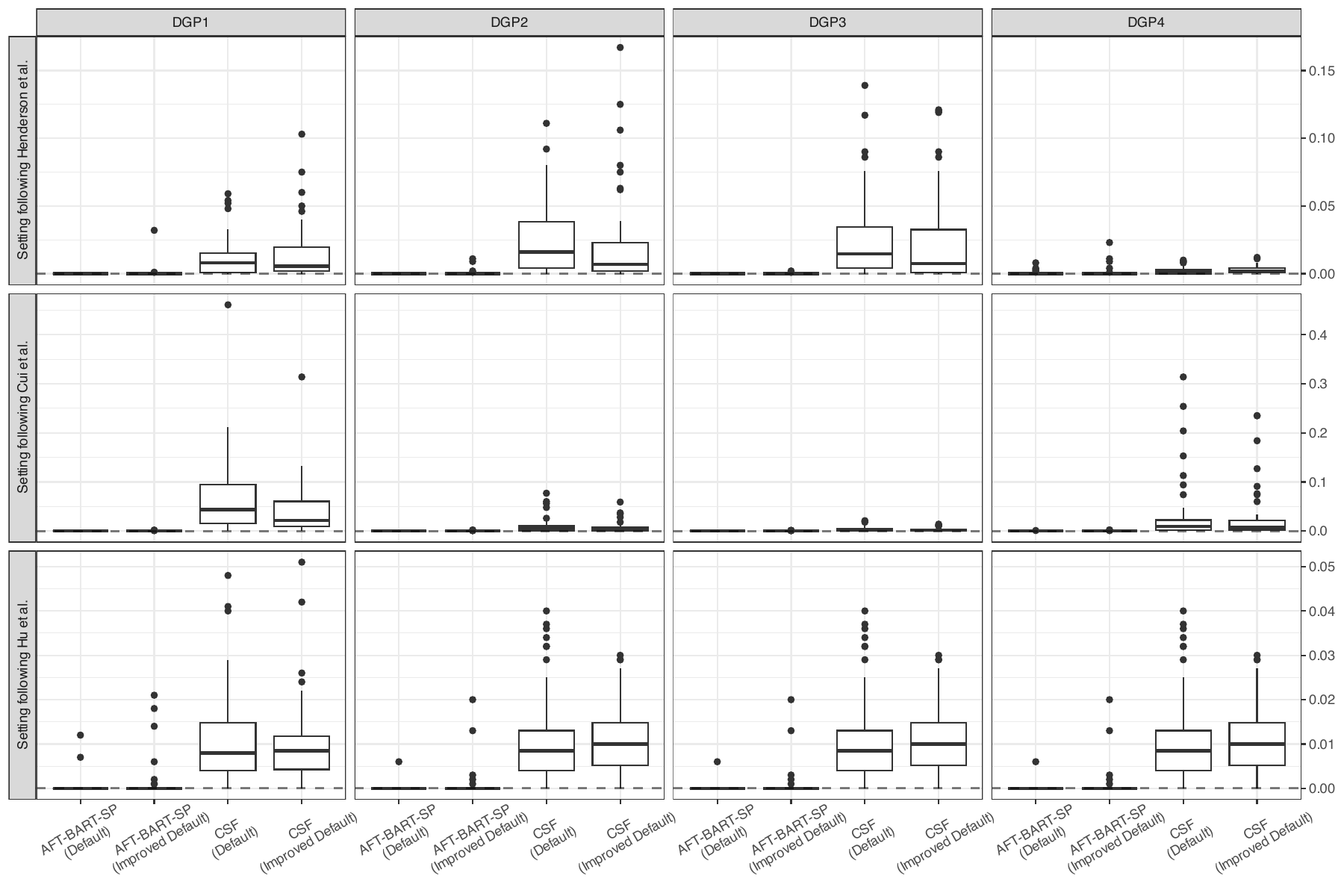}
\end{figure}

We modified each DGP from the Henderson, Cui, and Hu frameworks so that every individual had the same treatment effect (i.e., no variation in the ITE function). Apart from making the effect constant, all other features of the DGPs---covariate distributions, censoring processes, and sample sizes---remained unchanged. We then fitted both AFT-BART-SP and CSF exactly as before, using both the default and ``improved default'' hyperparameter settings.

To determine whether each method erroneously indicates individual-specific heterogeneity in the null setting, we applied separate criteria aligned with the inferential frameworks of AFT-BART-SP and CSF.
For an estimation using AFT-BART-SP, we used the approach in Henderson et al.\ (2020), examining the posterior probabilities of differential treatment effect
\[
D_i \;=\; P\bigl(\theta(\mathbf{x}_i) \;\geq\; \bar{\theta} \mid \mathbf{Y}, \bdelta \bigr),
\]
where
\(\bar{\theta} \;=\; n^{-1}\sum_{i=1}^{n}\theta(\mathbf{x}_i)\) 
denotes the average treatment effect and $(\mathbf{Y}, \bdelta)$ denotes all the observed survival outcomes. Importantly, the average treatment effect parameter $\bar{\theta}$ is computed for each posterior draw (rather than only from the posterior mean). For each individual \(i\), if \(D_i \leq 0.025\) or \(D_i \geq 0.975\), we interpret it as ``strong evidence'' that the ITE \(\theta(\mathbf{x}_i)\) differs from the overall treatment effect \(\bar{\theta}\). We then calculate the proportion of individuals who meet these criteria.

For estimation using CSF, we constructed a 95\% confidence interval for each individual’s \(\theta(\mathbf{x}_i)\) and checked whether it excludes the forest-estimated mean effect \(\bar{\theta}\). Specifically, if \(\bar{\theta}\) lies entirely outside the 95\% confidence interval for \(\theta(\mathbf{x}_i)\), we conclude that CSF identifies the \(i\)-th individual’s effect as significantly different from the population average.

Figure~4 summarizes the results for each DGP under the Henderson, Cui, and Hu frameworks. Under the default parameters, AFT-BART-SP almost never excluded the population-average treatment effect, consistently reporting a proportion of 0\% in most DGPs, and at most 0.1\% under the ``improved default'' setting. In contrast, CSF yielded higher exclusion proportions under both its default and improved settings—occasionally reaching 2--3\% for the Henderson and Hu frameworks and as high as 6--7\% in certain Cui scenarios.

\section{Discussion} 

In this study, we compared the performance of AFT-BART and CSF in quantifying the uncertainty of ITE estimates and other metrics for right-censored survival data. Our simulations were drawn from three distinct designs: the settings originally proposed for AFT-BART (\cite{henderson2020individualized}), those for CSF (\cite{cui2023}), and an additional ``neutral'' scenario (\cite{hu2021}). Across all these contexts, BART produced interval estimates whose empirical coverage was consistently closer to nominal levels than those of CSF, regardless of whether default or modified hyperparameters were used. This superior coverage performance underscores the potential benefits of a Bayesian modeling framework in right-censored contexts, where flexible and robust uncertainty quantification is essential.

Although coverage was our primary focus, we also evaluated other performance metrics (e.g., bias, RMSE, misclassification proportion, and calibration statistics) to gain a more complete understanding of how each method behaved across different data-generating processes. In certain scenarios—particularly some of those based on \cite{cui2023}—CSF’s bias or RMSE could be comparable to or even slightly lower than that of AFT-BART. However, these occasional advantages did not translate into better interval coverage for CSF, which generally fell short of the nominal 95\% level in our simulations. Moreover, AFT-BART’s advantage in coverage did not come at the expense of dramatically worse results on other metrics. Indeed, AFT-BART retained relatively low bias, displayed stable RMSE, and demonstrated more consistent calibration under a variety of parameter settings. These findings indicate that AFT-BART’s superior interval coverage is not merely an artifact of overly conservative intervals or a trade-off with accuracy; rather, it is indicative of the robust uncertainty quantification afforded by a Bayesian modeling framework.

An additional simulation scenario considered in our study was the “null” HTE case, where no true heterogeneity existed. In this setting, both BART and CSF indicated the absence of a meaningful treatment interaction on average, but BART’s posterior intervals often showed more conservative coverage. Such conservatism reduces the chance of false positives—an important advantage in clinical decision-making, where overestimating the presence of heterogeneity could potentially lead to suboptimal or harmful treatment allocations. Nevertheless, certain research or regulatory contexts may favor methods that are less conservative if the cost of false negatives is deemed more serious.

Beyond the direct comparison of AFT-BART and CSF, our study resonates with broader discussions in the literature about the advantages of Bayesian methods for HTE estimation. As noted by \cite{henderson2016bayesian} and  \cite{Segal2023-wm}, a Bayesian framework permits direct probability statements about treatment benefits, making it well-suited for individualized decision-making. The use of prior information allows for a transparent incorporation of existing knowledge or external data, which can be particularly valuable when dealing with rare events or complex censoring patterns. Caron’s recent work (\cite{Caron2022-an})  similarly illustrates how sophisticated Bayesian models can accommodate diverse data structures and large-scale complexities without sacrificing interpretability. Although Caron’s research primarily addresses network data, the general principle that Bayesian approaches can flexibly incorporate prior distributions and manage computational challenges applies to survival analysis settings as well. Taken together, these discussions support the notion that Bayesian methods are adaptable, conducive to principled inference, and highly relevant to clinical studies where the accurate quantification of individualized effects and their uncertainties is paramount.

Although our findings highlight the relative advantages of BART-based estimators, they also need to be interpreted with caution in light of the concerns raised by Efron in his comment to Breiman's discussion paper (\cite{Breiman2001-wo}). Efron emphasized that simulation-based comparisons can be prone to unintended biases, especially if the “old” benchmark methods are not optimized as diligently as newer or more sophisticated approaches. Our attempt to use simulation designs from previously published studies was intended to mitigate some of these issues, but it is still possible that other data-generating mechanisms, not explored here, could alter the comparative performance of the two methods. 
Furthermore, we examined only a subset of hyperparameter choices for both AFT-BART and CSF, even though we included both default and modified settings. It is possible that more extensive or alternative tuning strategies could improve the performance of CSF or even further enhance BART. 

Based on our simulation results and the broader discussions in the literature, our conclusion is that Bayesian methods—exemplified here by BART—are recommended for HTE estimation in right-censored survival data. While CSF remains a viable alternative, especially in large data settings, the advantages of a full Bayesian framework for individualized decision-making including direct probability statements about treatment benefits, flexible prior specification, and more direct quantification of uncertainty make BART-like methods particularly appealing. Future work should integrate more elaborate theoretical insights, additional simulation scenarios, and empirical applications to further illuminate the trade-offs between different estimators. Nonetheless, this study adds to the growing body of evidence favoring Bayesian approaches for precise and transparent estimation of heterogeneity in treatment effects.

\section{Acknowledgments} 
Dr. Varadhan was supported by the Cancer center core grant  NCI CCSG P30 CA006973.
Dr. Henderson was supported by the National Cancer Institutes of Health under Award Number P30 CA046592. The content is solely the responsibility of the authors and does not necessarily represent the official views of the National Institutes of Health.

\bibliographystyle{agsm}
\bibliography{SurvBART_refs}

%% APPENDIX ----------------------------------------------------------------
%\newpage
%\medskip
%\begin{center}
%{\large\bf APPENDIX}
%\end{center}

\appendix

\section{Comparisons of Semiparametric AFT-BART and Nonparametric AFT-BART}
We evaluated whether the key findings reported in this study would hold if we replaced the semiparametric version of the AFT-BART (AFT-BART-SP) with the nonparametric version (AFT-BART-NP) originally proposed by \cite{henderson2020individualized}. We adopted the semiparametric version in our main simulations primarily because it entails fewer computational complexities and can be more straightforward to tune for typical right-censored survival scenarios, making it more directly comparable to the structure of CSF. However, AFT-BART-NP relaxes certain assumptions about the error distribution in the survival model, potentially capturing more complex data patterns when heterogeneity is pronounced or when parametric assumptions are tenuous.

To confirm that our main results do not depend critically on the semiparametric specification, we repeated the neutral Hu's setting simulations using both AFT-BART-SP and AFT-BART-NP under two sets of hyperparameters: the default settings and an “improved default” that emerged from preliminary tuning. Figure A1 summarizes the performance measures, including bias, RMSE, misclassification proportion, the Hosmer–Lemeshow statistic, and coverage proportion, across four different data-generating processes (DGP1–DGP4). Overall, AFT-BART-NP performed comparably to AFT-BART-SP. In many cases, the nonparametric version showed either no meaningful difference in performance or a slight improvement, particularly in RMSE and coverage proportion. For instance, in DGP1, the improved default settings yielded coverage proportions of 0.930 for AFT-BART-SP versus 0.926 for AFT-BART-NP, and in DGP2, the slight advantage of AFT-BART-NP was reflected in marginally lower bias and RMSE. Although these differences were not universally large, the pattern that emerged suggests that allowing more flexible (i.e., nonparametric) error structures can be beneficial in certain data scenarios.

These findings thus reinforce the robustness of our main conclusions regarding BART-based methods. Whether one employs the semiparametric or nonparametric form of AFT-BART, coverage properties and overall predictive performance remain strong, supporting the idea that a Bayesian approach to modeling heterogeneous treatment effects in survival settings is both flexible and reliable. At the same time, they highlight that researchers who have a priori concerns about specific distributional assumptions may wish to opt for AFT-BART-NP, particularly if computational resources and data richness permit modeling a fully nonparametric error structure.

%% FIGURES ----------------------------------------------------------------
%\newpage
%\section*{Figures}

\begin{figure}[H]
        \textbf{Figure A1. Comparizon Between Semiparametric AFT-BART and Nonparametric AFT-BART Under Hu's Setting}\\
        \includegraphics[width=17cm]{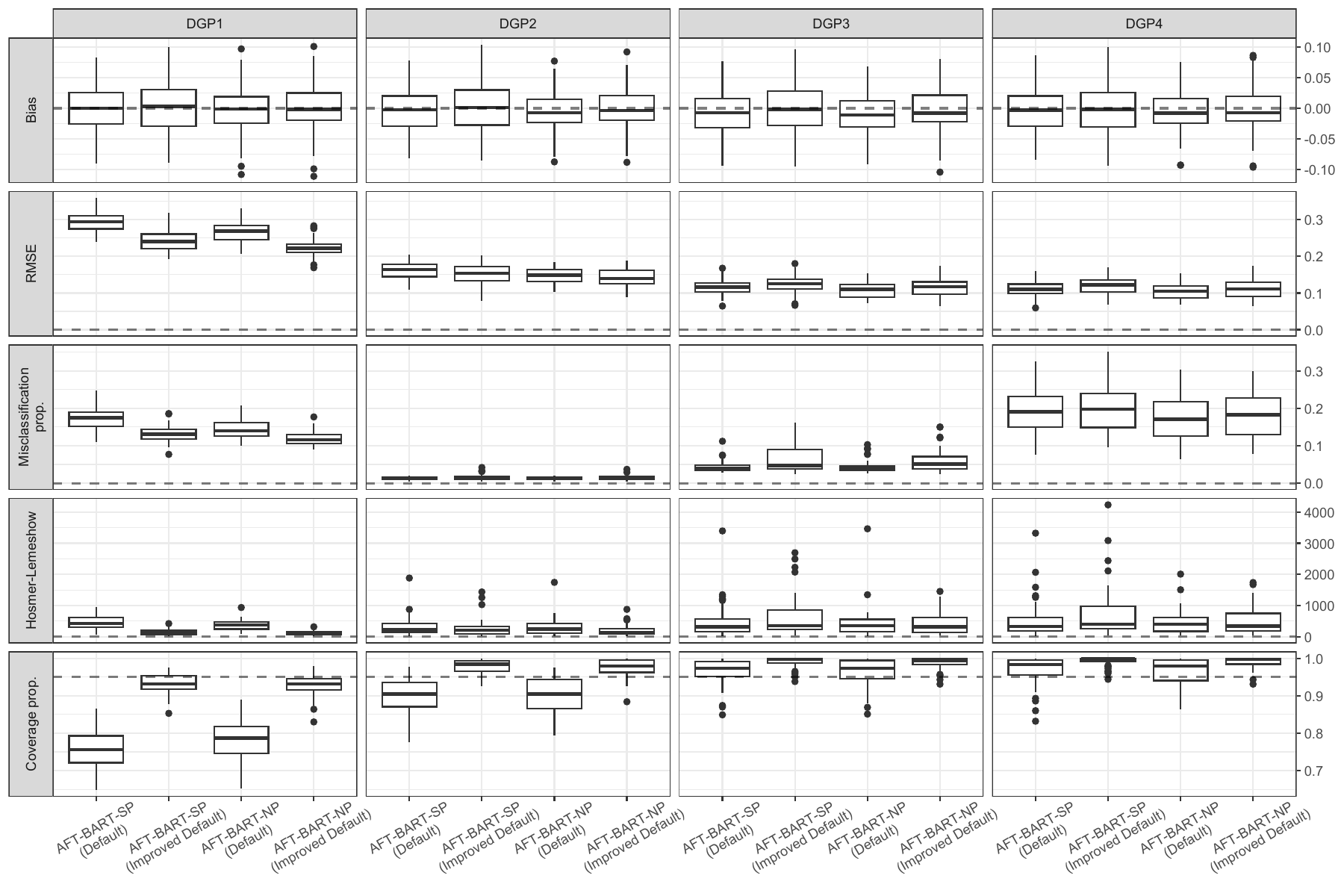}
\end{figure}

\end{document}